\documentclass[authoryear,5p,twocolumn]{elsarticle}
\def\aap{A\&A}%
\def\apjl{ApJ}%
\def\apjs{ApJS}%
\usepackage{epsfig}
\usepackage{color}

\usepackage{amssymb}
\usepackage{amsmath}
\usepackage{amsthm}
\usepackage{multirow}
\usepackage{twoopt}
\usepackage{graphicx}

\newcommand{\proton}{\ensuremath{^1}\textrm{H}}
\newcommand{\deut}{\ensuremath{^2}\textrm{H}}

\newcommand{\het}{\ensuremath{^3}\textrm{He}}
\newcommand{\hef}{\ensuremath{^4}\textrm{He}}

\newcommand*{\doi}[1]{\href{http://dx.doi.org/#1}{doi: #1}}

\usepackage[pdftex,             
            breaklinks=true,%
            colorlinks=true,%
            pdfauthor={Ghelfi et al.},%
            pdftitle={Template for manuscripts in Advances in Space Research}%
           ]{hyperref}

\makeatletter
\newcommandtwoopt{\citeads}[3][][]{\href{http://adsabs.harvard.edu/abs/#3}{\def\hyper@linkstart##1##2{}\let\hyper@linkend\@empty\citealp[#1][#2]{#3}}}
\newcommandtwoopt{\citepads}[3][][]{\href{http://adsabs.harvard.edu/abs/#3}{\def\hyper@linkstart##1##2{}\let\hyper@linkend\@empty\citep[#1][#2]{#3}}}
\newcommandtwoopt{\citetads}[3][][]{\href{http://adsabs.harvard.edu/abs/#3}{\def\hyper@linkstart##1##2{}\let\hyper@linkend\@empty\citet[#1][#2]{#3}}}
\newcommandtwoopt{\citealpads}[3][][]{\href{http://adsabs.harvard.edu/abs/#3}{\def\hyper@linkstart##1##2{}\let\hyper@linkend\@empty\citealp[#1][#2]{#3}}}
\newcommandtwoopt{\citealtads}[3][][]{\href{http://adsabs.harvard.edu/abs/#3}{\def\hyper@linkstart##1##2{}\let\hyper@linkend\@empty\citealt[#1][#2]{#3}}}
\newcommandtwoopt{\citeyearads}[3][][]{\href{http://adsabs.harvard.edu/abs/#3}{\def\hyper@linkstart##1##2{}\let\hyper@linkend\@empty\citeyear[#1][#2]{#3}}}
\newcommandtwoopt{\citeadsstar}[3][][]{\href{http://adsabs.harvard.edu/abs/#3}{\def\hyper@linkstart##1##2{}\let\hyper@linkend\@empty\citealp*[#1][#2]{#3}}}
\newcommandtwoopt{\citepadsstar}[3][][]{\href{http://adsabs.harvard.edu/abs/#3}{\def\hyper@linkstart##1##2{}\let\hyper@linkend\@empty\citep*[#1][#2]{#3}}}
\newcommandtwoopt{\citetadsstar}[3][][]{\href{http://adsabs.harvard.edu/abs/#3}{\def\hyper@linkstart##1##2{}\let\hyper@linkend\@empty\citet*[#1][#2]{#3}}}
\newcommandtwoopt{\citeyearadsstar}[3][][]{\href{http://adsabs.harvard.edu/abs/#3}{\def\hyper@linkstart##1##2{}\let\hyper@linkend\@empty\citeyear*[#1][#2]{#3}}}
\newcommandtwoopt{\citeauthoradsstar}[3][][]{\href{http://adsabs.harvard.edu/abs/#3}{\def\hyper@linkstart##1##2{}\let\hyper@linkend\@empty\citeauthor*[#1][#2]{#3}}}
\newcommandtwoopt{\citepthesis}[3][][]{\href{http://tel.archives-ouvertes.fr/docs/#3}{\def\hyper@linkstart##1##2{}\let\hyper@linkend\@empty\citep[#1][#2]{#3}}}
\newcommandtwoopt{\citetthesis}[3][][]{\href{http://tel.archives-ouvertes.fr/docs/#3}{\def\hyper@linkstart##1##2{}\let\hyper@linkend\@empty\citet[#1][#2]{#3}}}
\makeatother

\journal{Advances in Space Research}

\begin{document}

\begin{frontmatter}



\title{Neutron monitors and muon detectors for solar modulation studies: 2. $\phi$ time series}

\author[lpsc]{A. Ghelfi\corref{cor}}
\ead{ghelfi@lpsc.in2p3.fr}
\author[lpsc]{D. Maurin\corref{cor}}
\ead{dmaurin@lpsc.in2p3.fr}

\author[onera]{\\A. Cheminet}
\ead{Adrien.Cheminet@onera.fr}

\author[lpsc]{L. Derome}
\ead{derome@lpsc.in2p3.fr}

\author[onera]{G. Hubert}
\ead{Guillaume.Hubert@onera.fr}

\author[lpsc]{F. Melot}
\ead{melot@lpsc.in2p3.fr}

\cortext[cor]{Corresponding authors}

\address[lpsc]{LPSC, Universit\'e Grenoble-Alpes, CNRS/IN2P3,
      53 avenue des Martyrs, 38026 Grenoble, France}
\address[onera]{ONERA (French Aerospace Lab), 2 avenue Edouard Belin, 31055 Toulouse Cedex 4, France}

\begin{abstract}
The level of solar modulation at different times (related to the solar activity) is a central question of solar and galactic cosmic-ray physics. In the first paper of this series, we have established a correspondence between the uncertainties on ground-based detectors count rates and the parameter $\phi$ (modulation level in the force-field approximation) reconstructed from these count rates. In this second paper, we detail a procedure to obtain a reference $\phi$ time series from neutron monitor data. We show that we can have an unbiased and accurate $\phi$ reconstruction ($\Delta\phi/\phi\simeq 10\%$). We also discuss the potential of Bonner spheres spectrometers and muon detectors to provide $\phi$ time series. Two by-products of this calculation are updated $\phi$ values for the cosmic-ray database and a web interface to retrieve and plot $\phi$ from the 50's to today (\url{http://lpsc.in2p3.fr/crdb}).
\end{abstract}


\end{frontmatter}

\parindent=0.5 cm


\section{Introduction}

Measurements of top-of-atmosphere (TOA) cosmic-ray (CR) fluxes show a clear modulation related to solar activity \citepads{2013LRSP...10....1U}. The imprint of the 11-year solar cycle is present in secondary particles created in the Earth atmosphere \citepads{1974crvs.book.....D,2004ASSL..303.....D,2009crme.book.....D}, as seen in neutron monitor data \citepads{2000SSRv...93...11S}. Despite being an integral measurement (top-of-atmosphere fluxes folded by the atmosphere and instrument response), ground-based detectors have been providing monitoring of solar activity since the 50's, on a much finer timescale than balloon-borne and space experiments can achieve, even today.

In this work, we wish to provide a consistent description of modulation levels for cosmic-ray data. This is important in the context of galactic CR physics as clues on CR sources \citepads{2013A&ARv..21...70B} and constraints set on CR transport parameters \citepads{2007ARNPS..57..285S} are based on modulated CR data. Similarly, dark matter indirect detection \citepads{2012CRPhy..13..740L} involves low energy modulated antiproton and antideuteron fluxes. Unfortunately, the set of modulation levels provided for space or balloon-borne CR data (from the original publications) is not homogeneous and very patchy \citepads{2014A&A...569A..32M}: each value, when existing, is based on different assumptions regarding the IS spectrum (fitted to the experiment data, or resulting from different CR propagation models) and the modulation model (from force-field to sign-charge dependent drift models). This situation is inadequate and unsatisfactory.

Providing homogeneous modulation levels for past and present CR experiments or providing $\phi(t)$ time series are complementary tasks. In the context of the force-field approximation \citepads{1967ApJ...149L.115G,1968ApJ...154.1011G}, homogeneous monthly time series have been derived from NM data \citepads{1999CzJPh..49.1743U,2002SoPh..207..389U,2005JGRA..11012108U,2011JGRA..116.2104U} since July 1936. Note however, that many experiments operate on a shorter timescale, during which solar activity can significantly depart from the monthly average. This is especially true during solar maximum periods. Moreover, in the last years, the PAMELA\footnote{\url{http://pamela.roma2.infn.it}.} \citepads{2011Sci...332...69A,2013AdSpR..51..219A,2013ApJ...765...91A} and AMS\footnote{\url{http://www.ams02.org}.} \citepads{2015PhRvL.114q1103A,2015PhRvL.115u1101A} experiments provided high precision proton and helium fluxes. The latter can be used to improve the IS flux description \citepads{2016Ap&SS.361...48B,2015arXiv151108790C,2016A&A...591A..94G}, and in a second step the accuracy of $\phi$ time series.

Our approach is based on \citetads{2011JGRA..116.2104U}, with several differences. We build on our recent analysis of the uncertainties on $\phi$ reconstruction from ground-based detectors data \citepads[][hereafter Paper~I]{2015AdSpR..55..363M}. We also take advantage of our recent re-estimate of the interstellar (IS) proton and helium fluxes \citepads{2016A&A...591A..94G}. The robustness and consistency of $\phi$ time series from NM data (retrieved from the Neutron Monitor Data Base, NMDB\footnote{\url{http://www.nmdb.eu}.}) are validated against GCR data $\phi$ values and compared to other ground-based detector data.
 
The paper is organised as follows. In Sect.~\ref{sec:detectors}, we recall how IS spectra are modulated and folded by the yield function of ground-based detectors, whose data are used to reconstruct $\phi$ time-series. In Sect.~\ref{sec:sZsup2}, we discuss the enhancement factor to account for heavy CR contributions to count rates. In Sect.~\ref{sec:nm_norm}, the procedure to calculate the correction factor of a NM station is detailed. In Sect.~\ref{sec:time-series}, we calculate and compare $\phi$ time-series (and their uncertainties) as obtained from NM, GCR, Auger scaler, or neutron spectrometer data. We conclude in Sect.~\ref{sec:concl}. Along with the paper, we provide an online application to calculate, at any time in the past, $\phi$ values (for any time period) based on the methodology presented in this paper.

\section{Solar modulation and count rates from ground-based detectors}
\label{sec:detectors}

Count rate detector calculations and measurements, and their dependence on the environment (geomagnetic field, meteorological effects, yield function, etc.) are presented in the comprehensive monographs of \citetads{1974crvs.book.....D,2004ASSL..303.....D,2009crme.book.....D}. We briefly recall the ingredients of the calculation and our assumptions.

\subsection{From count rates to modulation parameters}
A ground-based detector ${\cal D}$ at $\vec{r}=(\varphi,\,\lambda,\,h)$ measures, at time $t$, a count rate $N^{\cal D}(\vec{r},t)$:
\begin{equation}
\label{eq:count-rate}
\!\!N^{\cal D}({\vec{r}},t) \!=\!\!\! \int_{0}^{\infty} \!\!\!\!{\cal T}(R,\vec{r}, t)
  \times \!\!\!\!\sum_{i={\rm CRs}} \!\!\!{\cal Y}^{\cal D}_i(R,h)\, \frac{dJ_{i}^{\rm TOA\!\!\!}}{dR}(R,t)  \;dR,
\end{equation}
where
\begin{itemize}
   \item ${\cal T}(R,\vec{r}, t)$ is the transmission function in the geomagnetic field. In practice, it is very often approximated by an effective vertical rigidity cutoff $R^{\rm eff}_c$ \citepads[see, e.g.,][for definitions]{1991NCimC..14..213C}, and this is the approach we follow here for simplicity. As discussed in Paper~I, using the  apparent cutoff rigidity or a sigmoid can lead to up to $50$~MV differences on the reconstructed $\phi$ values (stations with $R^{\rm eff}_c\lesssim 5$~GV are less sensitive to this effect).

   \item $\sum_{i={\rm CRs}}$ runs over all CR species. In practice, the He flux is rescaled by $(1+s_{Z>2})$ in order to sum over $i={\rm H,He}$ only. The factor $s_{Z>2}$ accounts for the contribution of species heavier than He \citepads{2003JGRA..108.1355W,2011JGRA..116.2104U,2015AdSpR..55..363M}, relying on the fact that the yield function for a CR nucleus of atomic mass A is A/4 times that of a CR helium \citepads{2011AdSpR..48...19M}. The analysis presented here updates the discussion of Paper~I, regarding $s_{Z>2}$ and its uncertainties.
     
   \item ${\cal Y}^{\cal D}_{i}(R,h)$ is the yield function, i.e. the detector response at altitude $h$ in count m$^2$~sr to a unit intensity of primary CR species $i$ at rigidity $R$. Yield functions are evaluated from the network of NMs \citepads{1989NCimC..12..173N,1990ICRC....7...96N,2012JGRA..11712103C} or from Monte Carlo simulations \citepads{1999ICRC....7..317C,2000SSRv...93..335C,2008ICRC....1..289F,2009JGRA..114.8104M,2013JGRA..118.2783M,chemi2013}. Our results are based on the Cheminet yield function (denoted C13) discussed in Paper~I, but we also discuss how using other parametrisations (gathered in App.~B of Paper~I) affect the results. We underline that all Monte Carlo-based calculations used in this study take into account the geometrical correction factor discussed in \citetads{2013JGRA..118.2783M}, which better fit the latitudinal survey count-rates (see Paper~I and \citealtads{2015JGRA..120.7172G}).
 
   \item $dJ_{i}^{\rm TOA}/dR$ is the top-of-atmosphere (TOA) modulated differential flux per rigidity interval $dR$ for the CR species $i$ in m$^{-2}$~s$^{-1}$~sr$^{-1}$~GV$^{-1}$. TOA fluxes are obtained from a modulation model (and its parameters) applied to IS fluxes.
   \begin{itemize} 
     \item Modulation model: in this study, we use the force-field approximation \citepads{1967ApJ...149L.115G,1968ApJ...154.1011G}, in which, for a given species $i$,
\begin{eqnarray}
\label{eq:forcefield}
  \!\!\!\!\!\!\!\!\frac{E_i^{\rm TOA}}{A}&\!\!\!=\!\!\!&\frac{E_i^{\rm IS}}{A} - \frac{|Z|}{A} \phi\;, \\
  \!\!\!\!\!\!\!\!J_i^{\rm TOA} \left( E_i^{\rm TOA} \right)&\!\!\!=\!\!\!&
  \left( \frac{p_i^{\rm TOA}}{p_i^{\rm IS}} \right)^{2} \times J_i^{\rm IS} \left( E_i^{\rm IS} \right). \nonumber
\end{eqnarray}
In the above expression and throughout the paper, for short, the differential flux per kinetic energy per nucleon interval is denoted
\begin{equation}
 J_i \equiv \frac{dJ_i}{dE_{k/n}} = \frac{A}{\beta Z} \frac{dJ_i}{dR}.
\end{equation}
This modulation model has only one free parameter, namely the modulation level $\phi$ (which should not be confused with $\Phi=|Z|/A\times \phi$).

   \item $J_{i}^{\rm IS}$ are the H and He IS fluxes used to calculate TOA fluxes. As discussed in Paper~I, the uncertainty on $J_{i}^{\rm IS}$ is one of the main source of uncertainty for $\phi$. As was also underlined in Paper~I, this uncertainty can be decreased taking advantage of recent high precision measurements. We rely below on the recent non-parametric determination of the IS H and He fluxes of \citetads{2016A&A...591A..94G}, which has a few percent uncertainty in the GeV/n to TeV/n range.
   \end{itemize}
\end{itemize}

These quantities, their uncertainties, and their impact on the determination of $\phi$ values were at the core of Paper~I (in particular, see their Table~9), to which we refer the reader for more details.

\subsection{Detector types}
We now briefly introduce the three types of ground-based detectors whose count rates will be used in this study (more details and references can again be found in Paper~I).

\paragraph{Neutron Monitors} Standardised NMs have been widely used across the world since the 50's \citepads{2000SSRv...93...11S}. They provide count rates with a one per minute frequency. The $n$-NM64 type consists of $n$ BF$_3$ proportional counter tubes surrounded by a cylindrical polyethylene moderator, inserted in a large volume of lead \citepads{1964CaJPh..42.2443H}. Most of the NM counts come from primary CRs in the range $1-500$~GV, and the secondary particles contributing to the rates are predominantly neutrons ($\sim 87\%$), but also protons ($\sim 8\%$), and $\mu^-$ ($\sim 5\%$).

\paragraph{Neutron spectrometers} Bonner Sphere Spectrometers (BSS) are a set of homogeneous polyethylene spheres with increasing diameters, each sphere hosting a high pressure $^3$He spherical proportional counter in its centre. Some of them include inner tungsten or lead shells in order to increase the response to neutrons above 20~MeV. BSS are only sensitive to neutrons, in the range $10^{-2}$~meV to GeV, with at best one spectrum per hour for high altitude stations. BSS were deployed at ground level and mountain altitudes to characterise the CR-induced neutron spectrum for dosimetry and microelectronics reliability purposes (\citealt{ruhm2009continuous}; \citealtads{2013ITNS...60.2418H}). 
The knowledge of the atmospheric radiations and their dynamics are essential issues in the evaluation of the Single Event Effects, the assessment of radiation risks in avionics/ground applications and the space environment (space weather). To study over a long and short period the dynamics of neutron spectrum from meV to GeV, neutron spectrometers are now operated simultaneously in three high-altitude stations in medium geomagnetic latitude and Antarctica environment: the first one was installed at the summit of the Pic-du-Midi in the French Pyr\'en\'ees (2885~m above sea level) in May 2011, the second at the summit of the Pico dos Dias in Brazil (1864~m asl), and the third one in the Concordia station (Antarctica 75°06'S, 123°20'E, 3233~m asl) since December 2015. 

In this analysis, we rely on the {\sc acropol}\footnote{High Altitude Cosmic Ray ONERA/Pic du Midi Observatory Laboratory.} BSS at the Pic du Midi (\citealtads{2012JInst...7C4007C,2012ITNS...59.1722C,2013ITNS...60.2411C,Cheminet2014RPD}). BSS are used in a NM mode, integrating over the neutron efficiency ${\cal E}_n^{\textrm{NM}}$ \citepads{2000SSRv...93..335C} of a NM times the BSS neutron fluence $\varphi_{n}^{\text{BSS}}$:
\begin{equation}
\label{eq:BSStoNM}
N_{X\,{\rm tubes}}^{\textrm{NM}}(\vec{r},t)=\frac{X}{6} \int _0^\infty {\cal E}_n^{\textrm{NM}}(T_n) \,\,\dot\varphi_{n}^{\text{BSS}} (T_n,\vec{r},t)\,\,dT_n\,.
\end{equation}
We recall that neutrons amount to most ($\sim 87\%$, see Paper~I) but not all of the total count rate in NMs.

\paragraph{Auger Scaler data} The Auger Surface Detector (SD) covers a total area of 3000~km$^2$, and this area is used to exposure determinations at the ultra high energy range. However, for low energy measurements 
as the scalers, the collection area is 16600~m$^2$, corresponding to 1660 water Cherenkov detectors (WCD) of 10~m$^2$ \citepads{Asorey2009,2011JInst...6.1003P,2015NIMPA.798..172B}. The threshold of the scaler mode is very low with a very high efficiency, providing a very good sensitivity to secondary particles. The WCD response indicates that the scalers are dominated by electromagnetic particles with a small contribution of muon \citep{2011ICRC...11..467A,2012AdSpR..49.1563D,Masias-Meza2015}. The Auger scaler data (corrected for pressure) are publicly available (15 min average) since 2005.\footnote{\url{http://auger.colostate.edu/ED/scaler.php}.} Given a proper modelling of the instrument response, these data could be used in principle to reconstruct $\phi$ time-series over this period. We will comment on this possibility in Sect.~\ref{sec:comparison}.

\section{Contribution of heavy CRs: weight factor $s_{Z>2}$}
\label{sec:sZsup2}
Cosmic ray species up to Fe give significant contributions to the count rates of ground-based detectors (see in particular Table~1 in Paper~I), and species heavier than He are generally accounted for as an effective enhancement $s_{Z>2}$ of the He flux in the calculation \citepads{2003JGRA..108.1355W,2011JGRA..116.2104U,2015AdSpR..55..363M}. Writing $N(R,t)=\int_{R_c}^{\infty}  {\cal I}(R,t)\;dR$ with ${\cal I}(R,t)\equiv\sum_{i={\rm CRs}}{\cal Y}_i(R,h) dJ_{i}^{\rm TOA}/dR$, then dropping implicit dependencies and recasting the sum in the integrand, the He flux enhancement factor $s_{Z>2}$ is the quantity that solves
\begin{equation}
\label{eq:integrand}
   {\cal I} \equiv \sum_i ({\cal Y}J)_i \approx {\cal Y}_{\proton{}}\times J_{\rm H} + {\cal Y}_{\hef{}}\times (1+s_{Z>2})\times J_{\rm He}.
\end{equation}
The $\approx$ sign stems from the fact that neither $J_{\rm H}$ nor $J_{\rm He}$ are pure \proton{} and \hef{} and that the contribution of nuclei heavier than He is not a perfect rescaled version of the He contribution (fluxes of various species are not scaled versions of one another).

\paragraph{$s_{Z>2}$ from spline-based fits}

The yield functions of nitrogen, oxygen, and iron scale to the helium yield function by their nucleonic number at the same energy per nucleon \citepads{2011AdSpR..48...19M,2013JGRA..118.2783M}. We assume here that it applies to all nuclei, ${\cal Y}_A(E_{k/n}) \approx (A/4) \times {\cal Y}_{\hef{}}(E_{k/n})$, and because CR nuclei have similar $A/Z\approx 2$, we use the same scaling for nuclei at the same rigidity, ${\cal Y}_A(R) \approx (A/4) \times {\cal Y}_{\hef{}}(R)$. With these approximations, the enhancement factor in Eq.~(\ref{eq:integrand}) becomes
\begin{equation}
\label{eq:corr_heavy}
s_{Z>2} \approx\left(\sum_{i=Z>2} A_i \frac{dJ^{\mathrm{TOA}}_i}{dR}\right) / \left(4\;\frac{dJ^{\mathrm{TOA}}_{\mathrm{He}}}{dR}\right)\,.
\end{equation}
As discussed in Paper~I, the scaling factor $s_{Z>2}$ is calculated for a given rigidity choice.
 
In practice, we retrieve all data for elements from Li to Ni up to 1 TV from the cosmic-ray database CRDB\footnote{\url{http://lpsc.in2p3.fr/crdb}.} \citepads{2014A&A...569A..32M}. All the IS fluxes $J_{\rm IS}$ are described by cubic splines (piece-wise functions defined by polynomials connecting at some knots), shown to better fit the data than standard single or double power-law functions \citepads{2016A&A...591A..94G}. We then perform a global fit to simultaneously constrain the spline parameters for all elements and the solar modulation levels $\phi$ for each data taking period. The fit is repeated several times, removing at each iteration inconsistent datasets. We checked that accounting for the isotopic abundances in $Z>2$ elements \citepads{2003ApJ...591.1220L} or assuming elemental fluxes are dominated by their most abundant isotope gives a negligible difference in the calculation ($\Delta s_{Z>2}/s_{Z>2}\approx0.2\%$).

At 30~GV, we obtain $s_{Z>2} = 0.445 \pm 0.005$ (propagating the errors from the IS flux determination). This value is quite sensitive to the rigidity choice, as we obtain 0.414 at 10~GV and 0.454 at 50 GV (rigidity range around which the contribution to count rates of CR fluxes is maximal, see Paper~I). In any case, the residual of the full contributions (from all $Z>2$ CRs) to the He scaling is lower or at the percent level of all contributions at all rigidities (see Fig.~1 of Paper~I). It is also insensitive to the modulation level taken (from 200~MV to 1500~MV). We thus have
\begin{equation}
\label{eq:s_value}
s_{Z>2} = 0.445 \pm 0.005~({\rm IS~fit}) \pm 0.03~({\rm scaling~approx.}).
\end{equation}
This result does not account for the yield function scaling approximation uncertainty, which is expected to be subdominant compared to ${\cal Y}_{\proton{}}$ and ${\cal Y}_{\hef{}}$ uncertainties (see below).

\paragraph{Bias from \het{} (in He) and impact on $\phi$}
\label{sec:bias_nm_toa}
Most of the CR H and He measurements do not achieve isotopic separation, so that the standard approach is to take He as \hef{} only. This disregards the fact that $\sim 20\%$ (peaking at $\sim 1$~GeV/n) of the measured flux is made of \het{}, which is differently modulated (not same $Z/A$) and gives rise to a different yield (not same $A$) than that of \hef{}. As discussed in \citetads{2016A&A...591A..94G}, the pure \hef{} assumption leads to an overshoot of $\sim 60$~MV in $\phi$ determined from CR data. For NMs, correctly accounting for the \het{} and \hef{} composition leads to a $+3.5\%$ increase of $s_{Z>2}$ compared to the pure \hef{} assumption (the few percent \deut{} in H has no effect on the result).\footnote{Accounting for \het{}, we have $s_{Z>2}=0.460$, which is close to 0.480 from Paper~I (using parametric fits for the IS fluxes) but differs from 0.428 \citepads{2011JGRA..116.2104U} who used different IS flux assumptions. Note that some confusion exists in Paper~I regarding the quoted $s_{Z>2}$ values: $0.480$ was obtained adding $Z>2$ CRs only but accounting for \het{} in He, whereas Eq.~(6) quoting $s_{Z>2}=0.611$ should have been written $s_{\deut{}+\het{}+Z>2}=0.611$.} However, we find that it has no impact on the reconstructed $\phi$ (see Sects.~\ref{sec:k_mean} and~\ref{sec:phi_nm}).

\section{Count rate absolute normalisation}
\label{sec:nm_norm}
Each detector requires an extra normalisation factor due to its specificities (local environment, electronics, detector). For a station $s$ and a count rate calculated with a yield function $y$, at any time $t$,
\begin{eqnarray}
\label{eq:nm_counts}
\displaystyle N^{\rm data}_{s}(t) \!\!\!\!&=&\!\!\!\! k^{\rm corr}_{s,\,y} \times N^{\rm calc}_{s,\,y}(t) \\
  &=& \!\!\!\! k^{\rm corr}_{s,\,y} \int_{R_c^s}^{\infty} \sum_{i={\rm CRs}} {\cal Y}^{y}_i(R,h)\, \frac{dJ_{i}^{\rm TOA\!\!\!}}{dR}(R,t)  \;dR.\nonumber
\end{eqnarray}
The normalisation is assumed to be independent of time once any change in the number of tubes over time is accounted for, as indicated in the station specifications (or identified from jumps in reconstructed $\phi$ time-series a posteriori).

\subsection{Correction factor $k^{\rm corr}_{s,\,y,\,\rm exp}$}
The correction factor is obtained equating the measured and calculated count rates at times for which $J^{\rm TOA}_{\rm H,\,He}$ fluxes are measured:
\begin{equation}
\label{eq:corr_factor}
k^{\rm corr}_{s,\,y,\,\rm exp} =
   \frac{\langle N^{\rm data}_{s}\rangle_{\rm exp}}{N^{\rm calc}_{s,\,y,\,\rm exp}(J^{\rm TOA}_{\rm H,\,He})},
\end{equation}
where $\langle N^{\rm data}_{s}\rangle_{\rm exp}$ denotes the count rate average over the data taking period of the CR experiment.

\begin{figure}[!t]
\begin{center}
\includegraphics[width=\columnwidth]{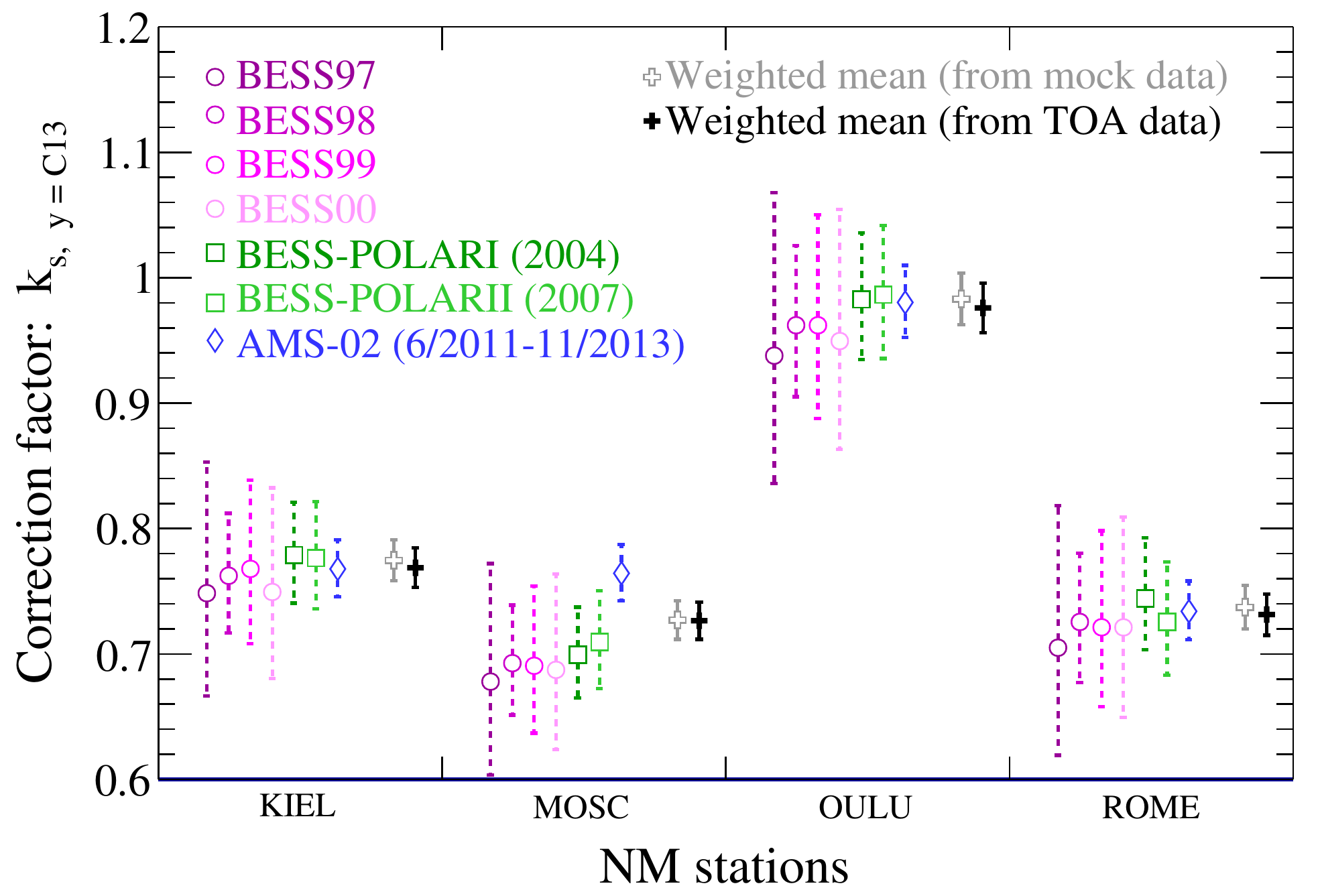}
\end{center}
\vspace{-0.25cm}
\caption{Correction factors Eq.~(\ref{eq:corr_factor}) based on the yield function C13 (Paper~I) using BESS97-00 \citepads{2007APh....28..154S}, BESS-PolarI \& II \citepads{2016ApJ...822...65A}, and AMS-02 \citepads{2015PhRvL.114q1103A,2015PhRvL.115u1101A} data taking periods. The error bars include TOA data errors and NM data statistical errors (negligible). The weighted average values and their uncertainties Eqs.~(\ref{eq:corr_mean}) and~(\ref{eq:Delta_corr_mean}) are computed from these data (empty `+' symbols) or mock data (filled `+' symbols) of the same experiments (see text for details).}
\label{norm_fact}
\end{figure}
Were the CR fluxes to be perfectly measured and the yield and transfer functions perfectly known, $k^{\rm corr}_{s,\,y,\,\rm exp}$ should not vary from one CR experiment to another. However, some inconsistencies are known to exist between CR datasets. For all the calculations below, we restrict ourselves to experiments that have measured both the TOA H and He fluxes (in a given data taking period), and apply the consistency-cut criterion discussed in \citetads{2016A&A...591A..94G}.\footnote{This leads to the exclusion of AMS-01 and PAMELA data (see \citealtads{2016A&A...591A..94G} for more details).} Because CR data have a limited energy coverage, each dataset was also completed above 200 GV by the IS flux from \citetads{2016A&A...591A..94G} to properly account for the small though significant contribution of this energy range to the total count rates in $N^{\rm calc}_{s,\,y}(\rm exp)$.

The correction factors per experiment $k^{\rm corr}_{s,\,y,\,\rm exp}$ are shown as symbols with dashed error bars for several stations in Fig.~\ref{norm_fact}. Our careful selection of TOA CR data leads to very stable values. The dashed segments correspond to $\Delta k^{\rm corr}_{s,\,y,\,\rm exp}$ obtained by propagating the uncertainties on the measured TOA fluxes. As expected, these uncertainties decrease in the more recent experiments (more accurate data). The slightly off AMS-02 correction factor for the Moscow station is at odd with the behaviour observed for most stations and could be an undocumented station setup modification (not related to the number of tubes as the change in the reconstructed $\phi$ values is gradual).

\subsection{Weighted average correction factor $\langle k^{\rm corr}_{s,\,y} \rangle_{\rm exp}$}
\label{sec:k_mean}
To obtain a single number for the correction factor and its uncertainty, we rely on the weighted average estimator \citep[e.g.,][]{lista2015statistical} over CR experiments:
\begin{eqnarray}
\label{eq:corr_mean}
\langle k^{\rm corr}_{s,\,y} \rangle_{\rm exp} & = &\displaystyle\frac{\left(\displaystyle \sum_{\rm exp}
   \displaystyle\frac{k^{\rm corr}_{s,\,y,\,\rm exp}}{\left(\Delta k^{\rm corr}_{s,\,y,\,\rm exp}\right)^2}\right)}{ \displaystyle \left(\sum_{\rm exp}\displaystyle\frac{1}{\left(\Delta k^{\rm corr}_{s,\,y,\,\rm exp}\right)^2}\right)}\,, \\
\label{eq:Delta_corr_mean}
\Delta\langle k^{\rm corr}_{s,\,y} \rangle_{\rm exp} & = & \left(\displaystyle\sum_{\rm exp}\displaystyle\frac{1}{\left(\Delta k^{\rm corr}_{s,\,y,\,\rm exp}\right)^2}\right)^{-1/2}\,.
\end{eqnarray}

The weighted average values are shown with a `+' black symbol with solid error bars. As a consistency check, the same weighted correction factors are calculated from simulated TOA datasets for all the above experiments, as generated from the IS fluxes and solar modulation levels taken from \citetads{2016A&A...591A..94G}: the latter values, that should be free of any residual inconsistencies in the real TOA datasets, are shown as empty `+' grey symbols in Fig.~\ref{norm_fact}. The difference observed between the two calculations is taken as a systematic uncertainty. This gives an overall uncertainty of
\begin{equation}
\label{eq:k_err_budget}
 \frac{\Delta\langle k^{\rm corr} \rangle_{\rm exp}}{\langle k^{\rm corr} \rangle_{\rm exp}} = 2.2\%,
\end{equation}
regardless of the station and yield function considered. Propagating the uncertainties related to the weight factor $s_{Z>2}$ (see Sect.~\ref{sec:sZsup2}) leads to $\Delta k/k<\pm 0.1\%$ (uncertainties from $J^{\rm IS}_{Z>2}$ fits), $\Delta k/k<\pm 0.6\%$ (approximation of a rigidity-independent scaling $s_{Z>2}$), and $\Delta k/k=-0.2\%$ (related to \het{} in He), which is sub-dominant in the total error budget.

\subsection{Impact of the yield function on $\langle k^{\rm corr}_{s,\,y} \rangle_{\rm exp}$}

\begin{figure}[!t]
\begin{center}
\includegraphics[width=\columnwidth]{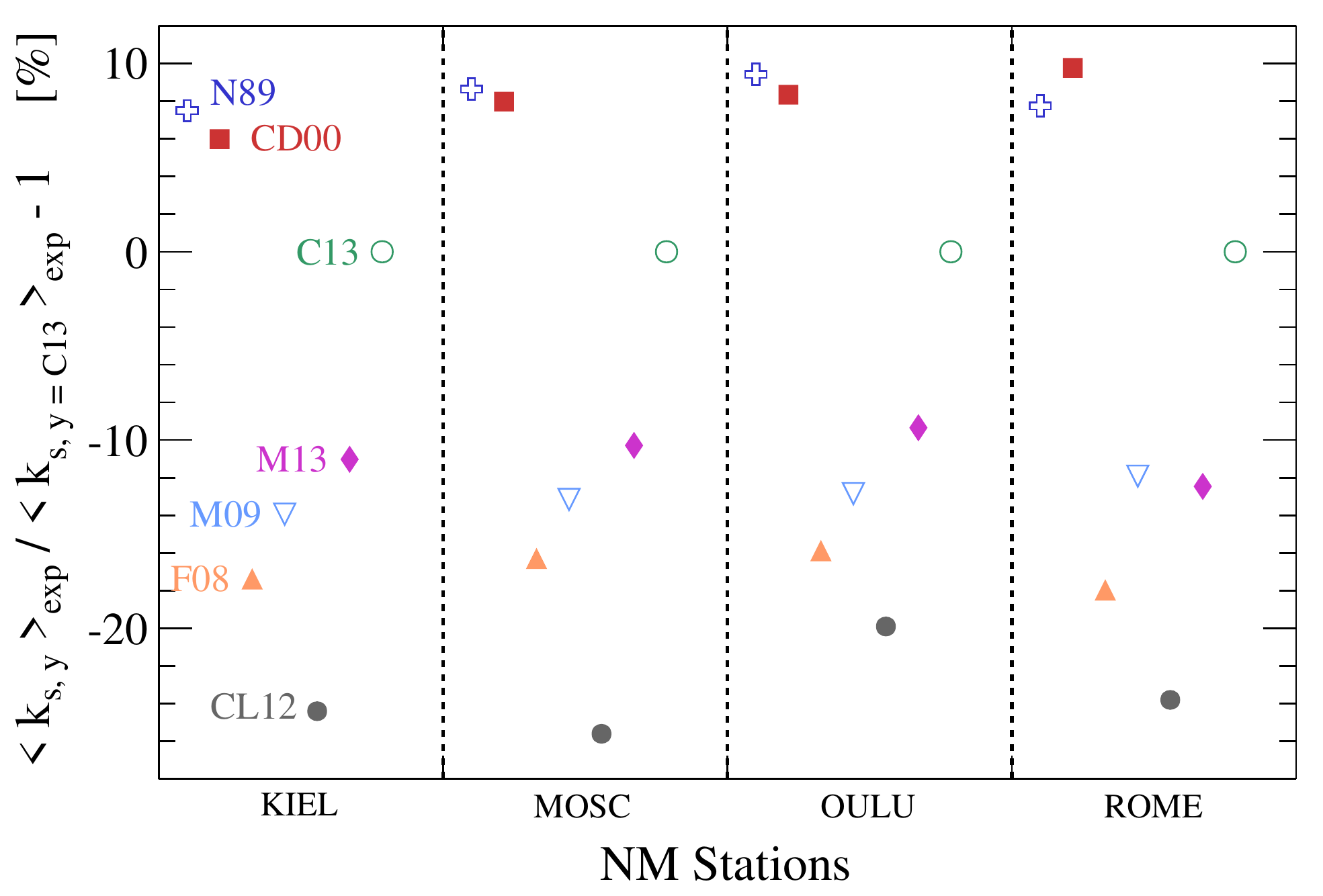}
\end{center}
\vspace{-0.25cm}
\caption{Relative difference (in percent) between the correction factors $\langle k^{\rm corr}_{s,\,y} \rangle_{\rm exp}$ defined in Eq.~(\ref{eq:corr_mean}) of different yield functions: N89 \citepads{1989NCimC..12..173N}, CD00 \citepads{2000SSRv...93..335C}, F08 \citepads{2008ICRC....1..289F}, M09 \citep{matthia_thesis}, CL12 \citepads{2012JGRA..11712103C}, M13 \citepads{2013JGRA..118.2783M}, and C13 (Paper~I) used as reference. See text for discussion.}
\label{fig:norm_fact_all}
\end{figure}

The previous results can be generalised to different yield functions. The latter have different normalisations and energy dependencies (see, e.g., Fig.~8 of Paper~I) and their scatter gives a fair indication---though probably too conservative---of the yield uncertainty. Shown in  Fig.~\ref{fig:norm_fact_all} are the factors obtained for six yield functions. As discussed in paper~I, Monte Carlo based yield functions (CD00, F08, M09, M13, C13) include \citetads{2013JGRA..118.2783M} geometrical correction factor, whereas NM-based yield functions do not (N89, CL12\footnote{All the results presented for this yield function are based on Eq.~(2) of \citetads{2012JGRA..11712103C} with $F_0 = 4.37 \cdot 10^{-4}$, $P_0=0.089$, $a=0.9$, $\gamma_1=0.748$ and $\gamma_2=61.3$, instead of the values reported in their Table~1 which contains misprints (Caballero-Lopez, private communication).}). Overall, there is a $\sim \pm 25\%$ spread in the correction factors. Different stations show similar results, with the same spread for the same yield functions.

\begin{table*}[!t]
\centering
\caption{Weighted mean averaged correction factors $\langle k^{\rm corr}_{s,\,y} \rangle_{\rm exp}$ for various stations (ordered by decreasing $R_c$) and yield functions: N89 \citepads{1989NCimC..12..173N}, CD00 \citepads{2000SSRv...93..335C}, F08 \citepads{2008ICRC....1..289F}, M09 \citep{matthia_thesis}, CL12 \citepads{2012JGRA..11712103C}, M13 \citepads{2013JGRA..118.2783M}, C13 (Paper~I). The relative uncertainty on these factors is $2.2\%$ (see Sect.~\ref{sec:k_mean}). For comparison purpose, we show in square brackets the values obtained from a similar analysis by \citetads{2011JGRA..116.2104U} and in curly brackets by \citetads{2015JGRA..120.7172G}. These authors use different TOA datasets for the normalisation. We have $\Delta k/k\approx 2.2\%$ from Eq.~(\ref{eq:k_err_budget}), whereas \citetads{2015JGRA..120.7172G} report $\Delta k/k\approx 0.1\%$.}
\label{tab:corr_factors}
\vspace{0.25cm}
\begin{tabular}{lcrllllllll}
\hline
Station      & $R_c$ [GV]& h [m]&&\multicolumn{7}{c}{$\langle k^{\rm corr}_{s,\,y} \rangle_{\rm exp}$}\\\cline{5-11}
                &      &        && {\em N89}&{\em ~~~~CD00}  &{\em F08}& {\em ~~~~M09}  &{\em CL12}& {\em ~~~~M13}    & {\em C13}  \\\hline
Almaty               & 6.69 &\!\!3340&&  $0.751$ & $0.712$        & $0.687$ & $0.570$        & $0.476$ &  $0.562$          & $0.861$ \\
Rome            & 6.27 &    60  &&  $0.787$ & $0.802$ [0.921]& $0.599$ & $0.643$ [0.597]& $0.540$ &  $0.639$ \{1.151\}& $0.731$ \\
Moscow          & 2.43 &   200  &&  $0.789$ & $0.784$        & $0.608$ & $0.631$        & $0.520$ &  $0.651$ \{1.241\}& $0.727$ \\
Kiel            & 2.36 &    54  &&  $0.828$ & $0.817$ [0.823]& $0.637$ & $0.663$ [0.548]& $0.570$ &  $0.686$ \{1.185\}& $0.770$ \\
Newark          & 2.02 &    50  &&  $0.906$ & $0.899$        & $0.697$ & $0.724$        & $0.624$ &  $0.750$ \{1.100\}& $0.852$ \\
Kerguelen       & 1.14 &    33  &&  $1.100$ & $1.090$ [0.990]& $0.848$ & $0.878$ [0.662]& $0.754$ &  $0.913$ \{0.971\}& $1.010$ \\
Oulu            & 0.78 &    15  &&  $1.070$ & $1.060$ [0.948]& $0.821$ & $0.850$ [0.634]& $0.743$ &  $0.885$ \{1.006\}& $0.963$ \\
McMurdo         & 0.30 &    48  &&  $1.320$ & $1.300$        & $1.010$ & $1.050$        & $0.909$ &  $1.090$ \{0.789\}& $1.220$ \\
Thule           & 0.30 &    26  &&  $1.210$ & $1.200$        & $0.935$ & $0.968$        & $0.834$ &  $1.010$          & $1.120$ \\
SouthPole       & 0.10 &\!\!2820&&  $1.010$ & $0.993$        & $0.898$ & $0.799$        & $0.701$ &  $0.830$          & $1.270$ \\
TerreAdelie\!\! & 0.00 &    45  &&  $1.130$ & $1.120$        & $0.869$ & $0.899$        & $0.789$ &  $0.934$          & $1.030$ \\
\hline
\end{tabular}
\end{table*}

We gather in Table~\ref{tab:corr_factors} the correction factors\footnote{$J^{\rm TOA}$ is the isotropic unidirectional differential intensity $I$ in (m$^2$~s~sr~(GeV/n))$^{-1}$. The omnidirectional or integrated differential intensity is given by $J_2=\int I d\Omega$, whereas the flux, often used as an input for atmospheric models, is $J_1=\int I \cos(\theta) d\Omega$ \citep[see Chapter~I of][]{Grieder:529397}. For an isotropic CR intensity, $J_1=2J_2$. Whereas M13 uses a weighting factor $\cos(\theta)$ for the simulated particles (Usoskin, private communication), we had to multiply by 1/2 all the other yield functions to obtain the correction factors gathered in Table~\ref{tab:corr_factors}.} for all NM stations (in NMDB) for which data are available during data taking periods of the CR experiments used in Sect.~\ref{sec:nm_norm}. This ensures that we have the most reliable normalisation for the count rates of these stations. The observed anti-correlation between the correction factor and the rigidity cutoff $R_c$ indicates that some of our approximations/ingredients (rigidity cut-off, yield function, etc.) introduce a bias. The bias slightly decreases when moving from N89 to C13, which is associated to an expected improvement on the yield function description. Our values are in fair agreement with those of \citetads{2011JGRA..116.2104U}, reported in the table in square brackets. However, they significantly differ from the results of \citetads{2015JGRA..120.7172G}, reported in the table in curly brackets. The latter analysis shows a correlation instead of an anti-correlation with $R_c$, a difference which is yet to be understood.

At this stage, we do not need to discuss further these differences. The exact value of the correction factor for each station is important in the context of fully understanding NM devices and their calibration. As investigated in the next section, the important question here is whether or not these different stations are able to provide similar $\phi$ time series once normalised.

\section{Time series: result and comparisons}
\label{sec:time-series}

We briefly summarise the procedure to obtain $\phi$ times-series and uncertainties from NM data: (i) calculate the contribution $s_{Z>2}$ to count rate of species $Z>2$ CRs relative to those from He (Sect.~\ref{sec:sZsup2}); (ii) from carefully selected CR TOA H and He data (and using $s_{Z>2}$), evaluate the correction factor $k^{\rm corr}$ for which calculated count rates match measured count rates for these TOA data taking periods (Sect.~\ref{sec:nm_norm}); (iii) given IS H and He fluxes \citepads{2016A&A...591A..94G}, $s_{Z>2}$, and  $k^{\rm corr}$, search for $\phi^{\rm NM}(t)$ that minimises\footnote{We use the  {\sc Minuit} package \citep{Minuit} from the {\sc Root} CERN libraries \url{https://root.cern.ch} \citepads{1997NIMPA.389...81B}.} the difference between $N^{\rm data}(t)$ and $k^{\rm corr} \times N^{\rm calc}$ in Eq.~(\ref{eq:nm_counts}): repeated at all times, this provides $\phi$ time-series from count rate time-series.

\subsection{Bias and accuracy of $\phi^{\rm NM}(t)$ time-series}
\label{sec:phi_nm}

\begin{figure}[!t]
\begin{center}
\includegraphics[width=\columnwidth]{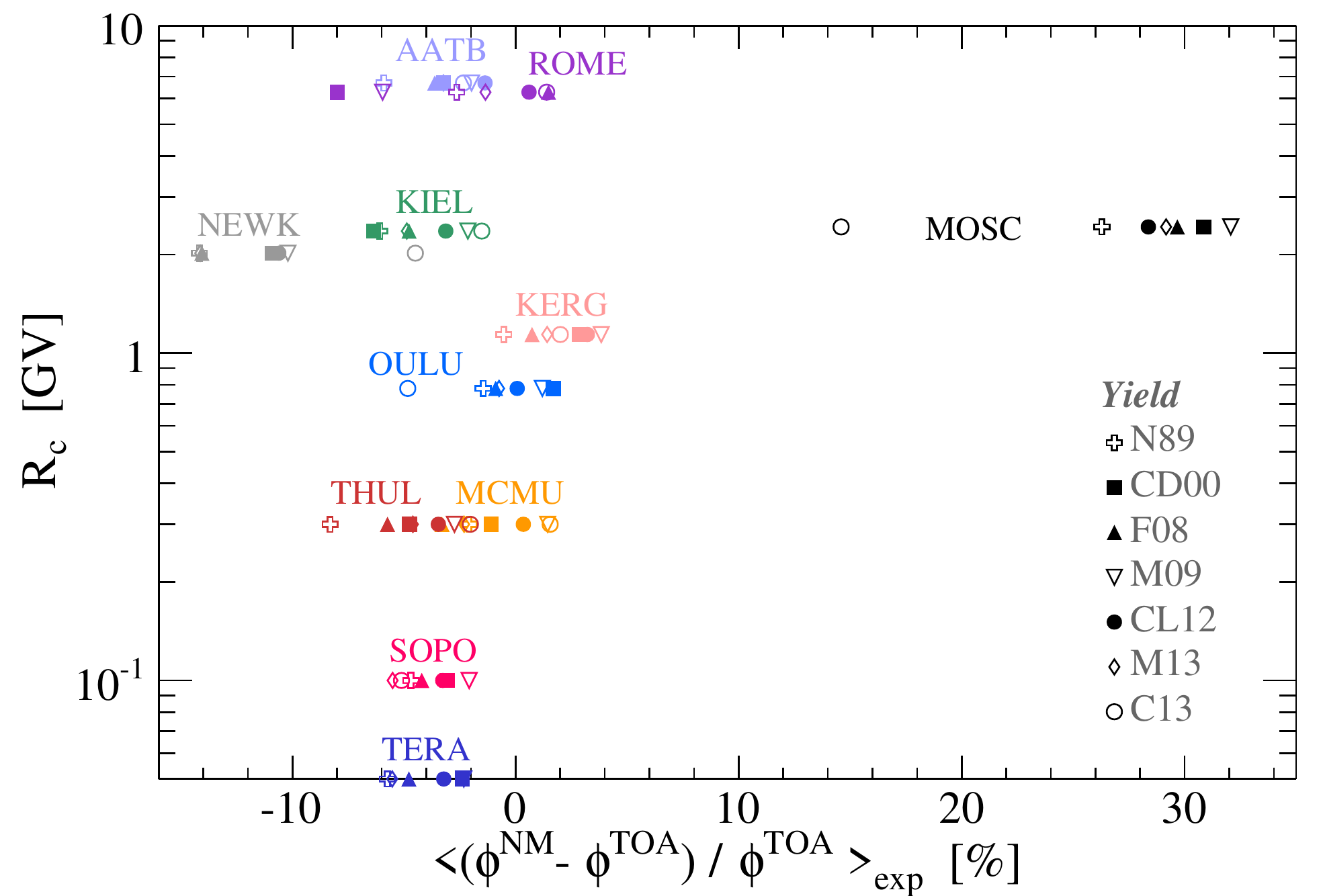}
\end{center}
\vspace{-0.25cm}
\caption{Average (over the selected periods of TOA measurements) of the relative difference between $\phi^{\rm NM}$ (calculated from NM count rates, this analysis) and $\phi^{\rm TOA}$ (calculated from TOA data, \citealtads{2016A&A...591A..94G}) for various yield functions (symbols) and stations given in Table~\ref{tab:corr_factors} (colours). Note that $R_c^{\rm TERA}=0$. }
\label{fig:distances}
\end{figure}

To evaluate how accurate $\phi^{\rm NM}(t)$ is, we repeat step~(ii) and~(iii) above for each NM station and yield function. We then compare the relative difference between these values (calculated for time periods of TOA experiments) and $\phi^{\rm TOA}_{\rm exp}$ obtained from the simultaneous determination of IS fluxes and modulation levels in \citetads{2016A&A...591A..94G}. To do so, we again use the CR dataset shown in Fig.~\ref{norm_fact}: the mean difference over all the associated time periods, i.e.
\begin{equation}
\label{eq:rel_diff_phi}
   \left\langle \frac{\Delta \phi^{\rm NM}}{\phi^{\rm NM}}\right\rangle_{\rm exp} \!\!\!\!= \frac{1}{n_{\rm exp}}\sum_{i=\rm exp}  \frac{\phi^{\rm NM}_{s,\,y}[t_i^{\rm start};t_i^{\rm stop}]-\phi^{\rm TOA}_i}{\phi^{\rm TOA}_i}\,
\end{equation}
is shown in Fig.~\ref{fig:distances} for all the yield functions and stations of Table~\ref{tab:corr_factors}. Whereas the correction factor varies by up to $\pm 25\%$ with the station, $\langle \Delta \phi^{\rm NM}/\phi^{\rm NM}\rangle_{\rm exp}\in [-10\%,\,0]$. The Moscow station which was an outlier of the correction factor analysis (pathological behaviour on the AMS-02 time period) is also an outlier here. Some stations fare slightly worse than others and can be disregarded; for instance, NEWK is centred around -12\% and ROME has a stronger dependence (larger spread) on the yield function.

The capability of the method to provide unbiased $\phi$ values from NM data is deduced from Fig.~\ref{fig:distances}. Most of the NM stations have an average deviation from TOA data of -5\%, but we recall that assuming He to be pure \hef{} positively biases  $\phi^{\rm TOA}$ by 60~MV \citepads{2016A&A...591A..94G} whereas it does not affect $\phi^{\rm NM}$ (Sect.~\ref{sec:bias_nm_toa}). Decreasing $\phi^{\rm TOA}$ by $\sim 5-10\%$ in Fig.~\ref{fig:distances} gives an excellent agreement with $\phi^{\rm NM}$, so that we conclude in a mostly unbiased determination of $\phi$ using NM data. At this stage, it is difficult to conclude about the origin of the remaining differences. It could be related to some systematics in the less recent CR measurements, position or time-dependent effects in the count rate calculation ($R_c$ approximation and/or $R_c(t)$), the IS flux and/or modulation model, or all of them. As discussed in Paper~I, small differences in any of these ingredients typically lead to $\sim 50$~MV changes for $\phi$ (see Table~9 in Paper~I). Properly taking into account the penumbra and non-vertical incident CRs \citepads[e.g.,][]{2008AdSpR..42..510D} as well as time dependencies of the rigidity cut-off \citepads[e.g.,][]{2009AdSpR..44.1107S} seems to be the next step to refine this study, but it goes beyond the scope of this paper.

The total error budget on $\phi^{\rm NM}$ reconstruction is obtained by combining all sources of errors tracked and combined at each step of the calculation:
\begin{itemize}
  \item  $\langle\sigma_{\phi^{\rm NM}}/\phi^{\rm NM}\rangle_{y} \approx \pm 6\%$ is the scatter obtained from the use of different yield functions (see Fig.~\ref{fig:distances}). It is taken to be the typical uncertainty on $\phi$ from our incomplete knowledge of the yield.

  \item $\langle\Delta\phi^{\rm NM}/\phi^{\rm NM}\rangle_{k^{\rm corr}} \approx \pm 11\%$ is the average difference (over the whole reconstructed time-series) between value reconstructed from $k^{\rm corr}$ and $k^{\rm corr}_{\rm min,max} = k^{\rm corr}\pm \Delta k^{\rm corr}$ from Eq.~(\ref{eq:k_err_budget}). Ongoing high precision measurements of the H and He fluxes with AMS-02 could decrease this number.

  \item $\Delta\phi_{J^{\rm IS}}^{\rm NM} \approx \pm 25$~MV is obtained by propagating the H and He IS flux uncertainties given in \citepads{2016A&A...591A..94G}.\footnote{Quoting either the relative or the absolute uncertainty relates to the fact that we found $\Delta\phi^{\rm NM}_{J^{\rm IS}}$ to be independent of the modulation level, whereas for the two first items, it is the case of the relative uncertainty (see also Paper~I that links $\Delta\phi/\phi$ to $\Delta N/N$).}
\end{itemize} 

For illustration purpose, we show in the top panel of Fig.~\ref{fig:phi_nm_selected} modulation levels $\phi^{\rm NM}(t)$ reconstructed from the Kerguelen station compared to $\phi^{\rm TOA}_{\rm exp}$ (and $\Delta\phi^{\rm TOA}_{\rm exp}$) calculated in \citetads{2016A&A...591A..94G}. The shaded grey area corresponds to $\Delta\phi^{\rm NM}$ obtained by quadratically combining the errors discussed above.

\subsection{Comparison to other types of ground-based detectors}
\label{sec:comparison}
\begin{table}[!t]
\centering
\caption{Rigidity cut-off, altitude, and data taking periods for the BSS \citepads{2013JGRA..118.7488C} and Auger scaler \citepads{2011JInst...6.1003P}. The last column presents the correction factor for the BSS and the relative change on the correction factor applied to Auger scaler in the corresponding time interval (next-to-last column). See text for discussion.}
\label{tab:corr_factors2}
\vspace{0.25cm}
\begin{tabular}{ccccc}
\hline
Station         &\!\!$R_c$ [GV]\!\!        & \!h [m]\!             &  Dates     &\!$\langle k^{\rm corr} \rangle_{\rm exp}$\!\!\!\\\hline
\!\!\!Pic-du-midi\! & \multirow{2}{*}{5.6} &\multirow{2}{*}{\!2885\!}& 30/05/2011 & \multirow{2}{*}{0.782$^\ddagger$} \\
(BSS)           &                      &                       & 14/02/2016 &                        \\[3mm]
                & \multirow{4}{*}{9.5} & \multirow{4}{*}{\!1400\!}& 20/09/2005$^\star$\!\!\!& \multirow{2}{*}{} \\
  Malarg\"{u}e  &                      &                       & 11/07/2007 & \multirow{2}{*}{(-2\%)$^\dagger$\!\!\!} \\
 \!\!(Auger scaler)\!\! &              &                       & 20/07/2010 & \multirow{2}{*}{(-2\%)} \\
                &                      &                       & 18/08/2015 &                        \\
\hline
\end{tabular}
\\
{\footnotesize $^\ddagger$ BSS used in NM mode only accounts for 87\% of the particles detected by a NM.\\
$^\star$ Data from 01/03 to 20/09 are based on a different WCD threshold \citepads{2011JInst...6.1003P} and are discarded.\\
$^\dagger$This correction was identified a posteriori in the reconstructed $\phi^{\rm scaler}(t)$, in agreement with the description given at \url{http://auger.colostate.edu/ED/scaler.php?spec=1}.}
\end{table}

\begin{figure}[!t]
\begin{center}
\includegraphics[width=\columnwidth]{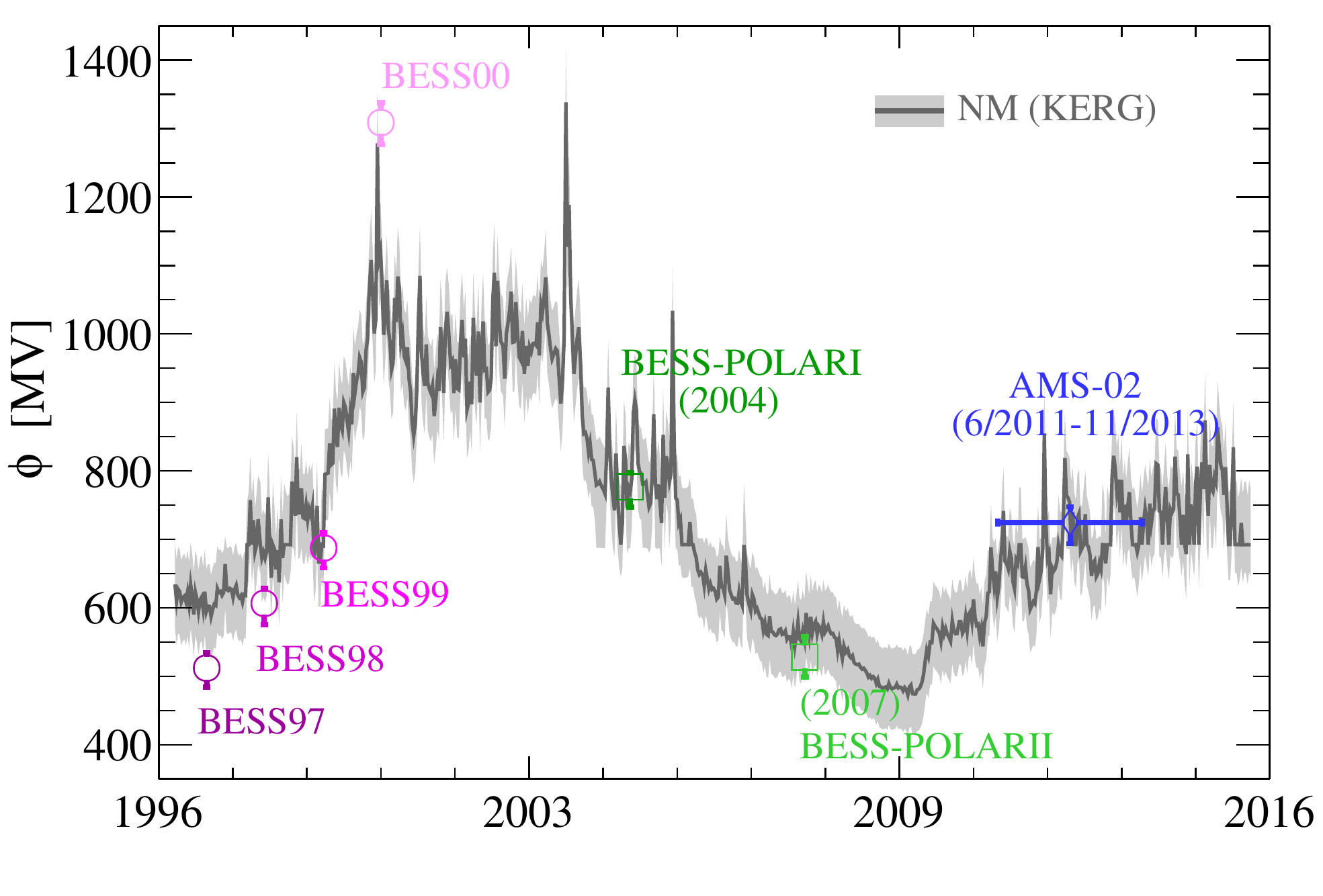}
\includegraphics[width=\columnwidth]{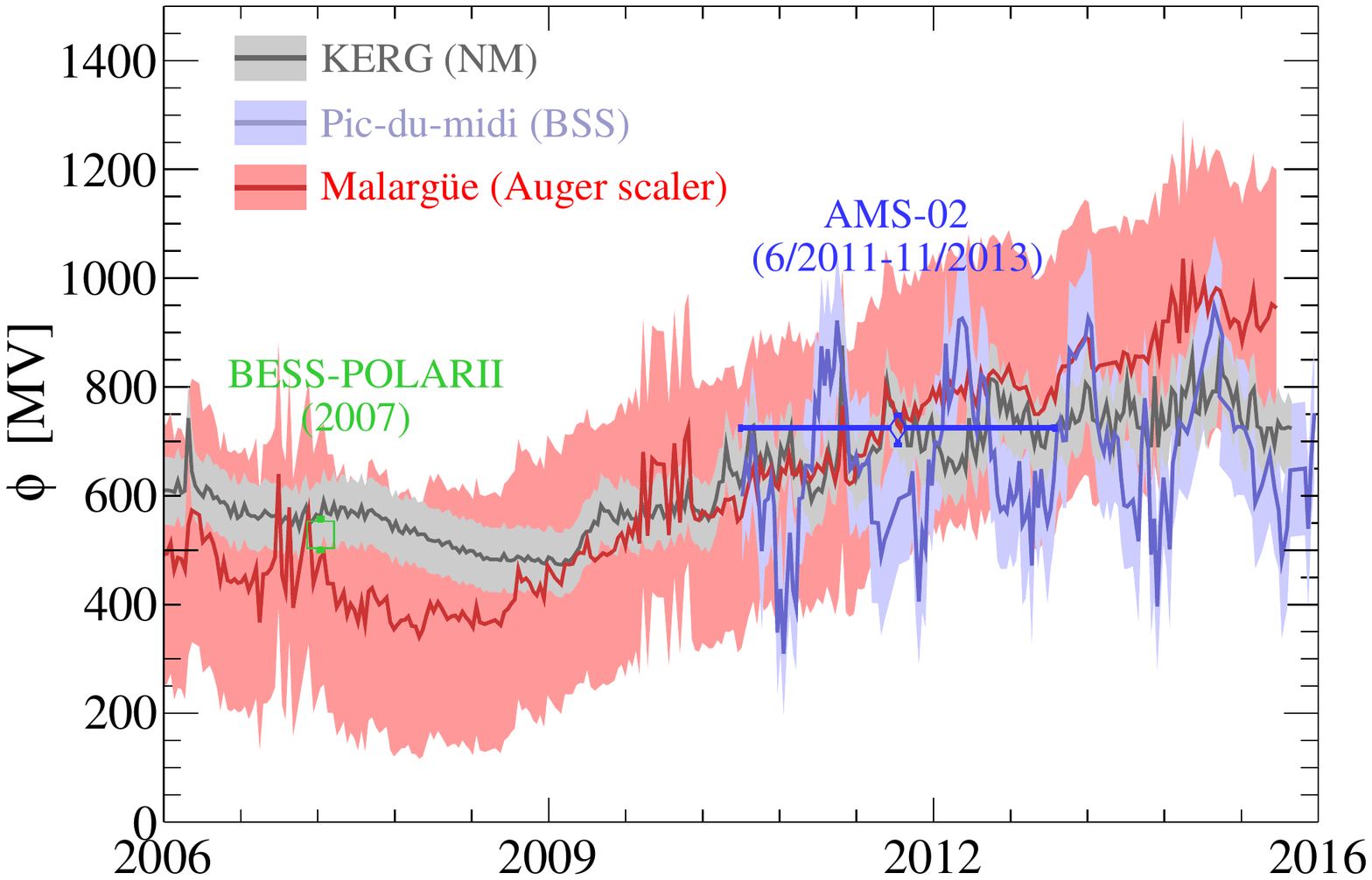}
\end{center}
\vspace{-0.6cm}
\caption{{\em Top panel:} ten-days average $\langle\phi^{\rm NM}\rangle$ time-series (solid line) and uncertainties (shaded area) displayed along with $\phi^{\rm TOA}$ \citepads{2016A&A...591A..94G} for illustration. We underline that $\phi^{\rm NM}$ calculated on the exact BESS97 time interval is much lower than the 10-days average and in full agreement with $\phi^{\rm TOA}$. {\em Bottom panel:} comparison of ten-days average $\langle\phi^{\rm NM}\rangle$ (grey), $\langle\phi^{\rm scaler}\rangle$ (red), and $\langle\phi^{\rm BSS}\rangle$ (blue) time-series. The symbols show the CR TOA data available to calculate the correction factor for BSS and Auger scaler data. See text for discussion.}
\label{fig:phi_nm_selected}
\end{figure}

We repeat the above steps to extract $\phi^{\rm BSS}(t)$ and $\phi^{\rm scaler}(t)$ time-series from the Pic-du-Midi BSS data \citepads{2013JGRA..118.7488C} and Auger scaler data \citepads{2011JInst...6.1003P} respectively. Table~\ref{tab:corr_factors2} indicates the rigidity cut-off and altitude for these detectors, as well as their data taking periods (available as of the writing of this paper). As illustrated in Fig.~\ref{fig:phi_nm_selected}, very few of our selected CR data measurements overlap with these data, namely AMS-02 only for BSS, and AMS-02 and BESS-PolarII for Auger scaler.

\paragraph{BSS: $k^{\rm corr}$, $\phi^{\rm BSS}(t)$, and uncertainties} BSS are sensitive to neutrons, which represent 87\% of the total count rates seen by a NM. A perfectly well calibrated BSS should require a correction factor of 0.87, whereas we find 0.782 (see Table~\ref{tab:corr_factors2}). This shows that BSS devices are as well calibrated but as sensitive to their environment as NM devices. We also have $\Delta\phi^{\rm BSS}(t)\approx\Delta\phi^{\rm NM}(t)$ (see bottom panel of Fig.~\ref{fig:phi_nm_selected}) because of the similar origin for the uncertainties: yield function scatter $(\pm6\%)$, correction factor ($\pm 11\%$) and IS flux ($\pm 25$~MV) uncertainties. However, relative to $\phi^{\rm NM}(t)$, $\phi^{\rm BSS}(t)$ shows a strong yearly variation. As discussed and estimated in Paper~I, an increase of the snow coverage for the Pic-du-midi station is responsible for a $-7\%$ drop of count rates (see Table~9 of Paper~I), leading to a 230~MV (resp. 460~MV) increase for the reference modulation level of 500~MV (resp. 1000~MV). This is typically the amplitude observed on the plot. This effect strongly biases $\phi^{\rm BSS}$ time-series and needs to be corrected for in order for BSS to be used for this purpose. Conversely, BSS data could help calibrating the snow coverage effect in NM data for station suffering from snow falls.

\paragraph{Auger scaler vs $\mu$-like detectors} The public Auger scaler data, when accounting for the WCD response are mostly sensitive to electromagnetic secondary particles \citepads{2011ICRC...11..467A,2012AdSpR..49.1563D}. The Pierre Auger collaboration also recently presented data from the so-called histogram mode in which vertical muons become dominant \citep{Masias-Meza2015}. Waiting for these data, our analysis is based on the scaler mode, assuming nonetheless a $\mu$ yield function (Paper~I) for the calculation. This provides a worst case reconstruction of $\phi^{\rm scaler}(t)$ on which the use of the histogram mode data could improve in the future. This also illustrates the expected uncertainties for `real' $\mu$-like detectors, which is the main goal of this section. Given these limitations, we note that the agreement observed in Fig.~\ref{fig:phi_nm_selected} between $\phi^{\rm scaler}(t)$ and $\phi^{\rm NM}(t)$ is already very encouraging.\footnote{We are aware of some efficiency effects of the detectors that are relevant for long term studies \citep{Masias-Meza2015}. At the time of preparation of this manuscript, long term efficiency corrections of the scaler rates were not available for the data publicly released by the Auger Collaboration. For a preliminary description of this correction and its effects, see \citet{Masias-Meza2015}. The Auger Collaboration is working to steer this study towards a journal publication, following which the public data will be updated.} The shaded area shows the uncertainty band calculated from the use of a $\mu$ yield function in the analysis. We recall that compared to NMs, $\mu$ detectors are sensitive to higher energy primary CRs (typically 100 GeV/n, see Fig.~10 in Paper~I) that are less sensitive to solar modulation. The relative count rate changes are smaller for $\mu$ detectors than for NMs. Consequently, whereas the correction factor uncertainty $\Delta k^{\rm corr}/k^{\rm corr}=3.6\%$ (from AMS-02 TOA data uncertainty only) is similar to that for NMs (mostly dominated by AMS-02, see Fig.~\ref{norm_fact}), the uncertainty on $\phi$ is much larger for $\mu$ ($\Delta\phi^{\mu}=310$~MV) than for NMs ($\Delta\phi^{\rm NM}/\phi^{\rm NM}=11\%$). The remaining uncertainties estimated in this paper are subdominant: the muon yield function is well known (see Paper~I) and the uncertainty from the IS flux is sightly smaller ($\pm18$~MV) than for NMs ($\pm25$~MV)---the IS flux is better constrained \citepads{2016A&A...591A..94G} in the energy range relevant for $\mu$ detectors. However, we underline that the temperature effects~\citepads{2011APh....34..401D}, if not corrected for, can lead to $\Delta\phi^\mu\lesssim 300$~MV (see Table~9 in Paper~I).

\subsection{Comparison between CR data, different stations, and previous calculations}

\begin{figure*}[!t]
\begin{center}
\includegraphics[width=\textwidth]{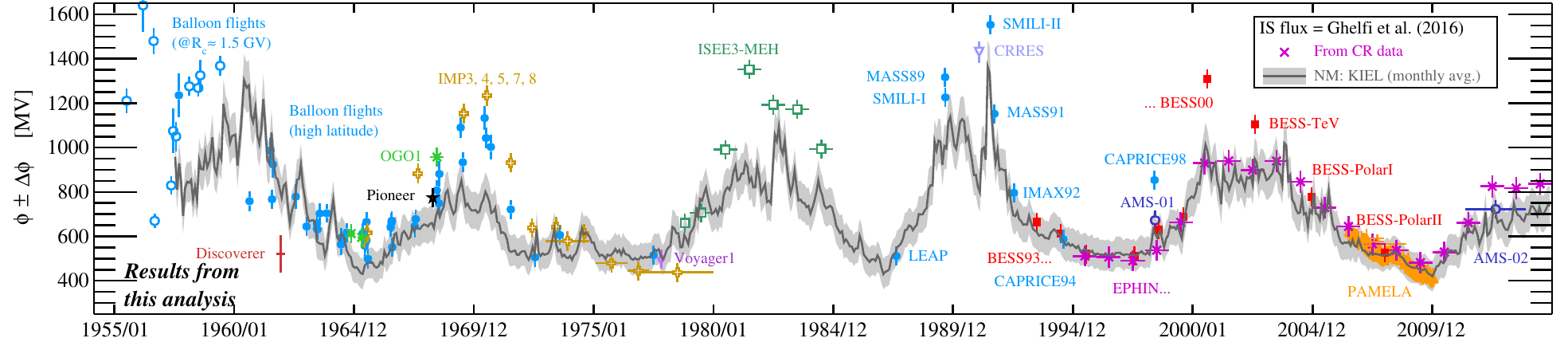}
\includegraphics[width=\textwidth]{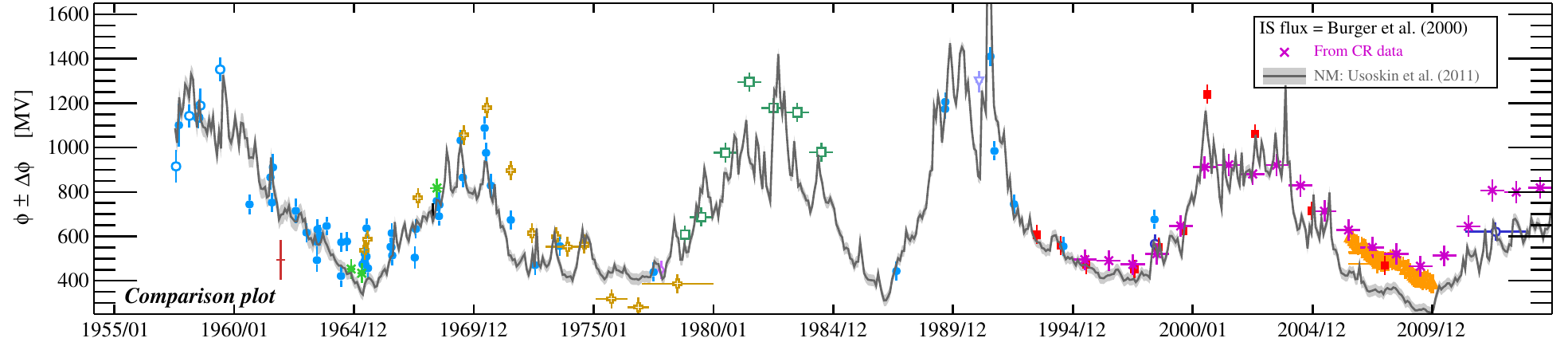}
\includegraphics[width=\textwidth]{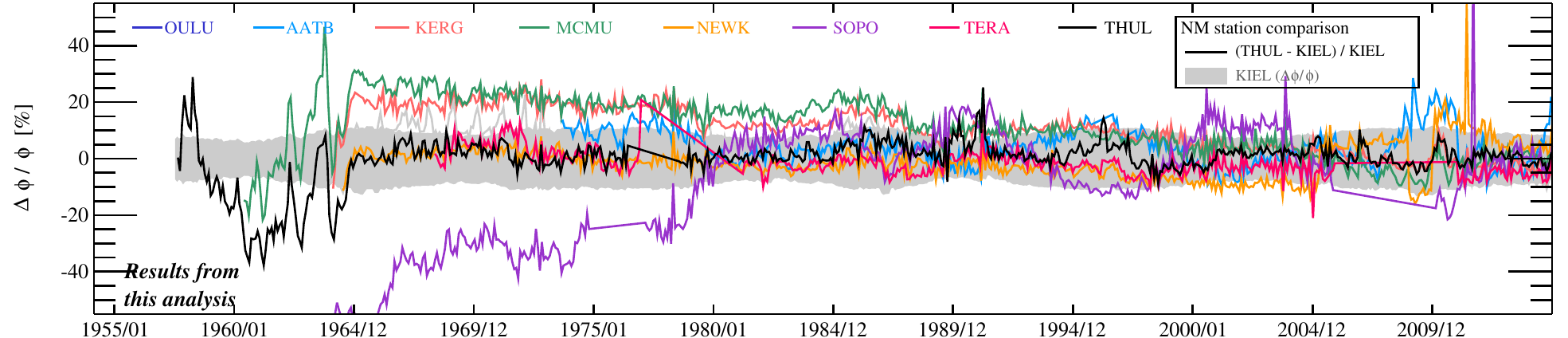}
\includegraphics[width=\textwidth]{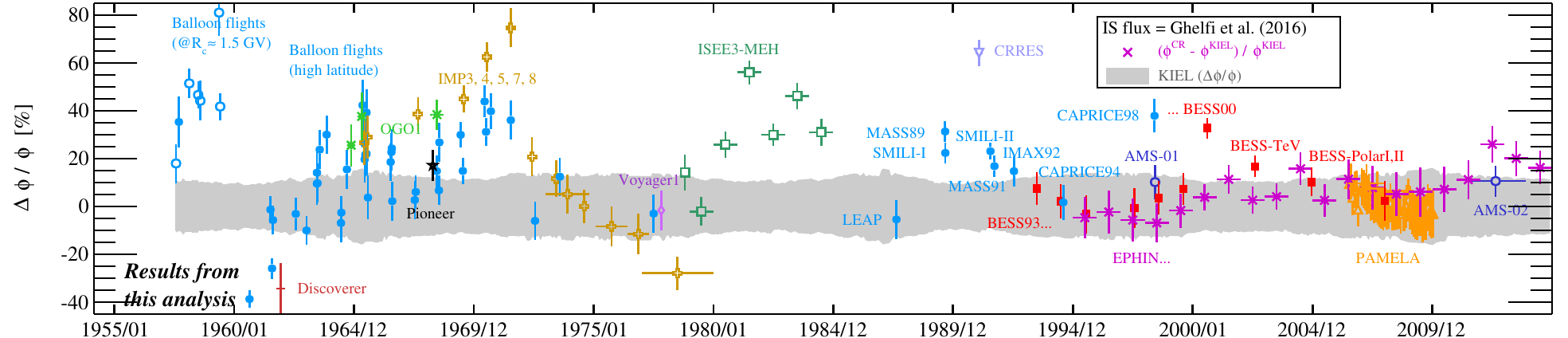}
\includegraphics[width=\textwidth]{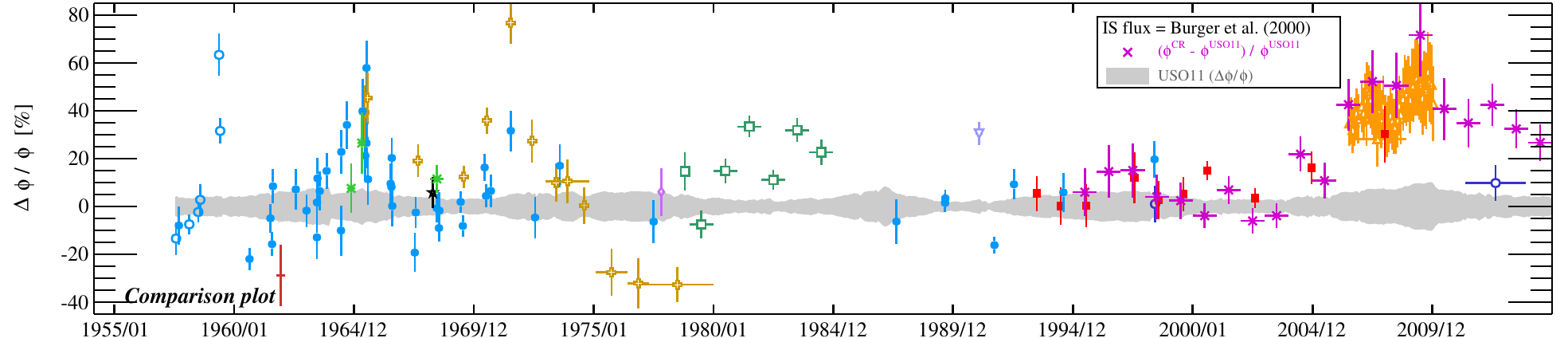}
\end{center}
\vspace{-0.2cm}
\caption{{\em Two top (bottom) panels:} (residuals of) $\phi^{\rm TOA}$ reconstructed from proton and helium CR fluxes, and monthly average $\langle\phi^{\rm NM}\rangle(t)$ time-series from NM data. The first (fourth) panel shows the results of this analysis based on the IS flux of \citetads{2016A&A...591A..94G} to calculate both $\phi^{\rm TOA}$ and $\phi^{\rm NM}$, while the second (last) panel is based on the IS flux hypothesis of \citetads{2000JGR..10527447B}, showing the corresponding NM time-series directly taken from \citetads{2011JGRA..116.2104U}|denoted USO11. The references for CR data are given in \ref{app:refs}. We underline that several experiments in the 50's took place at locations with 1.5~GV rigidity cut-off (empty circles), so that the reconstructed $\phi^{\rm TOA}$ may be biased for these data.
{\em Central panel:} residuals between NM calculations from various stations and KIEL station (used as a reference). In all the panels, the grey shaded area shows the 1$\sigma$ uncertainty estimated on $\phi^{\rm NM}$, from the full propagation of errors described in this work (panels 1, 3 and 4) or from \citealtads{2011JGRA..116.2104U} analysis (panels
2 and 5). In all panels (but panel 3), the 1$\sigma$ error bars on $\phi^{\rm TOA}$ are calculated accounted for IS flux uncertainties (see \citealtads{2016A&A...591A..94G}). See text for discussion.}
\label{fig:phi_all}
\end{figure*}

CR TOA data and NM data date back to the 50's, so that the stability and robustness of $\phi$ time series can be tested between various stations and detectors, comparing $\phi^{\rm NM}(t)$ and $\phi^{\rm TOA}$ values. It is also interesting to compare our long-term NM time-series to the one published by \citetads{2011JGRA..116.2104U}. This is shown in Fig.~\ref{fig:phi_all}, where our final results (panels labelled `Results from this analysis') are discussed and compared to \citetads{2011JGRA..116.2104U} results  (panels labelled `Comparison plot').

\paragraph{Comparison of $\phi^{\rm NM}$ and $\phi^{\rm TOA}$ values} The top panel shows the NM modulation level time series (monthly average) reconstructed from 1953 to 2015 for the KIEL station (solid line); the shaded area corresponds to the $1\sigma$ uncertainties obtained from the propagation of all uncertainties discussed in previous sections. Using the same IS flux hypothesis \citepads{2016A&A...591A..94G} as for the calculation of $\phi^{\rm NM}$, we fit $\phi^{\rm TOA}$ (symbols) to H (or \proton{}) and He (or \hef{}) TOA CR data retrieved from CRDB \citepads{2014A&A...569A..32M}. The error bars on all the points are calculated from propagating the uncertainty on both the CR data points and the IS fluxes. A fair agreement between the two sets of values is seen over the whole time period, be it for data from balloon-borne or satellite experiment, from fits to He data only (in some of the early experiments) or to p and He data. 

We recall that a perfect match is not expected since most of the TOA data have a data taking period of a few hours (during which large solar variations can happen) whereas, for the sake of legibility in this panel, the NM-reconstructed values are calculated on a monthly average. To be more quantitative, the fourth panel shows the residuals of the difference with $\phi^{\rm NM}(t)$ calculated on the appropriate time interval for NMs, i.e. an interval matching that of the CR data taking period. There is a trend for larger differences during strong solar activity (large $\phi$) periods than during low solar activity. This may partly be due to a limitation of the force-field approximation at low energy \citepads[e.g.,][]{2004JGRA..109.1101C} but also, as suggested by the improving agreement over time, from underestimated or systematic uncertainties in older CR data. From 1994, a typical 20\% difference remains between the two sets of values, which is almost completely accounted for (at the $1\sigma$ level) by the uncertainties in both reconstruction (grey area for NMs, and error bars for TOA CR data).

\paragraph{Comparison of $\phi^{\rm NM}$ from various stations}
The third panel shows a comparison of the reconstructed $\langle\phi^{\rm NM}\rangle(t)$ time-series (monthly average) from various NMs over 60 years (broken lines correspond to periods without data for the stations). Whereas our `calibration' procedure is based on CR data from the 1997-2015 period, the agreement between stations (at very different rigidity cut-off, see Fig.~\ref{fig:distances}) is very good down to the starting date of these devices. The residual of some stations (w.r.t. KIEL) is correlated with the solar activity (e.g., THUL and SOPO), indicating that choosing these stations would probably slightly improve the agreement between $\phi^{\rm TOA}$ and $\phi^{\rm NM}$ values. In any case, most of the stations are within 1$\sigma$ (grey band) of the reference (arbitrarily chosen to be KIEL): SOPO shows more variation, and some trend to depart from other stations is seen when moving back in time for KERG and TERA. This may be related to the long time evolution of the rigidity cut-off \citepads{2008ICRC....1..733S,2008ICRC....1..737S,2009AdSpR..44.1107S}, estimated to lead to a factor of 2 change in $\phi$ over 50 years ($\sim 13$~MV per year at most, see 
\citealtads{2015AdSpR..55..363M}). As shown in Fig.~3 of \citepads{2008ICRC....1..737S}, the maximum variation is located around the Atlantic ocean and the South Atlantic Anomaly. In particular, THUL is in the northern hemisphere, close to zones of increase of the rigidity cut-off, whereas SOPO is at the south pole, closer to regions of decreasing variation. An increase (decrease) of $R_c$ corresponds to a decrease (increase) of $\phi$
\citepads{2015AdSpR..55..363M}, which would move the time-series for THUL and SOPO in the desired direction  (see third panel of Fig.~\ref{fig:phi_all}). This is of course a very naive interpretation of a difference that could have another origin (e.g., a setup change in the station). Given the agreement and precision that is now reached between the stations, it would be interesting to estimate the variation of $R_c$ over the last 70 years for all the stations. Also, although we have only shown a couple of station, our analysis could easily be extended to other stations. This is left for future investigations.

\paragraph{Comparison with \citetads{2011JGRA..116.2104U} calculation} A reference calculation in the field is the time-series of \citetads{2011JGRA..116.2104U}, based on a different IS flux hypothesis; namely \citetads{2000JGR..10527447B}. We report their monthly average values\footnote{\url{http://cosmicrays.oulu.fi/phi/Phi_mon.txt}.} $\langle\phi^{\rm NM}\rangle(t)$ in the second panel of Fig.~\ref{fig:phi_all} (solid line and shaded area). Compared to our calculation (top panel), \citetads{2011JGRA..116.2104U} have a larger amplitude of variation: for solar minimum (maximum) periods, they obtain smaller (larger) modulation levels. Also, \citetads{2011JGRA..116.2104U} calculation is based on a weighted mean over several NM stations and yield functions, leading to uncertainties that are smaller than the one we have ($\sim 5\%$ vs $\sim 10\%$). In the same panel, we also recalculate $\phi^{\rm TOA}$ values from the same CR TOA dataset as in the top panel, but using \citetads{2000JGR..10527447B} IS flux. This choice leads to $\lesssim 200$~MV lower modulation levels for several sets of data (OGO1, SMILI-II, CRRES, BESS, etc.) compared to the top panel values obtained with \citetads{2016A&A...591A..94G} IS flux. This is related to the fact that different experiments cover different energy ranges and that the differences between \citetads{2000JGR..10527447B} and \citetads{2016A&A...591A..94G} fluxes are energy dependent. A comparison of $\langle\phi^{\rm NM}\rangle(t)$ to $\phi^{\rm TOA}$ shows a better agreement for \citetads{2011JGRA..116.2104U} analysis than for ours before the 90's. In particular, a surprisingly good agreement is obtained before 1964 (second panel), whereas we obtain a very different behaviour (top panel). The third panel shows that using THUL and MCMU instead of KIEL in our analysis would provide a behaviour closer to the one observed by \citetads{2011JGRA..116.2104U} in this period. However, for the present period (after 2004), $\langle\phi^{\rm USO11}\rangle(t)$ overshoots. For a more quantitative view, the bottom panel shows the residuals of the difference between TOA and NM calculations. A comparison of the next to last and last panels shows that where the TOA CR data precision is best and most reliable, our calculation is 
more successful. The seemingly better agreement at earlier periods for \citetads{2011JGRA..116.2104U} suggests that any effect that would impact differently the low and high energy range of the TOA spectrum could probably improve the agreement between NM and pre-90's low-energy CR TOA data (ISEE, IMP, balloons). This is exactly the features more evolved modulation models provide, and this would be interesting to investigate in a future study. On the other hand, one cannot exclude that systematics in the CR data are the main reason for the remaining differences.

\section{Conclusions}
\label{sec:concl}

We have revisited and extended the analysis of \citetads{2011JGRA..116.2104U} to refine the calculation of $\phi$ time series (and uncertainties) from three type of ground-based detectors (NM, BSS and $\mu$-like detectors):
\begin{itemize}
   \item Our analysis benefits from the improvements made on the determination of the H and He IS fluxes and their uncertainties \citepads{2016A&A...591A..94G}. The associated error $\Delta\phi_{J^{\rm IS}}$ is estimated to be no more than $\pm25$~MV for NMs and BSS, and $\pm18$~MV for $\mu$-like detectors, to be compared to the at-the-time conservative 200~MV of Paper~I.

   \item A common assumption to calculate count rates is to fold the H and He TOA fluxes by the yield function, accounting for $Z>2$ CRs as an enhancement factor $s_{Z>2}$ for He. Following the minute approach of \citetads{2016A&A...591A..94G} to extract $Z>2$ IS fluxes, we find $s_{Z>2}=0445\pm 0.03$. This uncertainty comes from the assumption of a rigidity-independent enhancement, which at present provides a negligible contribution to the total error budget of $\phi^{\rm NM,\,BSS,\,\mu}(t)$. Improvements in the calculation of $\phi^{\rm NM}$ will however require to use an energy-dependent scaling to keep this uncertainty subdominant.
   
   \item The uncertainty from the yield function is estimated from the scatter obtained when using different parametrisations, leading to $(\sigma_\phi/\phi)_y\approx 6\%$ for NMs and BSS (negligible for $\mu$). Actually, most of the yield function uncertainty is absorbed in the correction factor $k^{\rm corr}$ that must be estimated for each detector (to account for environment effects). The spread in count rates from the scatter of the yield function, taken at face value, leads to a 25\% scatter on $\phi$ (as estimated in Paper~I), but gives the above 6\% spread on $\phi^{\rm NM,\,BSS}$ when all the calculation steps are carried out.
   
   \item A key step is the calculation of $k^{\rm corr}$, which is tackled by the use of carefully selected TOA CR data on which to normalise the detector. This allows us to reduce the uncertainty to $\Delta k^{\rm corr}/k^{\rm corr}=\pm2.2\%$ ($\pm3.6\%$ for $\mu$), leading to a $\pm50$~MV uncertainty for NM and BSS, and $\pm310$~MV for $\mu$. This is the dominant source of uncertainty, by far for $\mu$-like detectors, but it could be decreased as more TOA measurements become available.

\end{itemize}

This analysis shows that with an improved IS flux description, $\phi$ values extracted from NM data are now in agreement with those extracted from TOA data|a similar conclusion was recently reached by \citetads{2015AdSpR..55.2940U} analysing a Forbush decrease. The two sets of data are complementary: NM count rate data have a good time sampling and a stable setup over time, which are useful properties to reconstruct robust $\phi$ time series; CR TOA data, though scarcer and sometimes suffering from systematics, provide differential fluxes with increasing precision (as new instruments and techniques are used) that help to properly calibrate IS fluxes and $\phi$ time series.

Further improvements of $\phi^{\rm NM}(t)$ time-series would require to take properly into account the Earth geomagnetic field and its long term evolution (for a better description of the transmission function). Improving on the yield function parametrisation and taking advantage of ongoing TOA measurements to further reduce the IS flux uncertainties is also desired. For the other detector types, BSS data show promising potential for intercalibration with NM data, in particular their complementarity to study snow falls effects on the neutron spectrum and count rates. Modulation time-series from $\mu$ detectors suffer from larger uncertainties than those from NM (due to the precision of TOA data used to calibrate their efficiency), but they also are complementary as they are not sensitive to the same uncertainties. In this respect, future Auger data from their histogram mode ($\mu$-like counter) are awaited to further investigate this complementarity.

To conclude, we refer the interested reader to \url{http://lpsc.in2p3.fr/crdb}, where we provide an online tool to extract $\phi^{\rm NM}(t)$ time series, and/or the average $\langle\phi^{\rm NM}\rangle_{\Delta t}$ on a given time interval, and/or modulated fluxes, based on this analysis. The new $\phi^{\rm NM}(t)$ values are also used to provide homogeneous sets of values for all CR data in CRDB.

\section*{Acknowledgements}
We warmly thank C. B\'erat and H.~Asorey, S.~Dasso, and P.~Ghia (Auger cosmo-geophysics task) for their help with Auger scaler data and very useful suggestions. We thank C.~Combet for a careful reading of the paper.  D.~M. warmly thanks Veronica Bindi, Peggy Shea, Don Smart, Christian Steigies, and Ilya Usoskin for illuminating discussions which took place at the workshop {\em ``Solar Energetic Particles (SEP), Solar Modulation and Space Radiation: New Opportunities in the AMS-02 Era''}. We acknowledge the NMDB database (\url{www.nmdb.eu}), founded under the European Union's FP7 programme (contract No. 213007) for providing data. This work has been supported by the ``Investissements d'avenir, Labex ENIGMASS" and by the French ANR, Project DMAstro-LHC, ANR-12-BS05-0006.

\appendix
\section{References for CR data}
\label{app:refs}
Due to a lack of space in the caption of Fig.~\ref{fig:phi_all}, the list of references associated to the CR data shown is presented in this appendix:
\begin{itemize}

   \item Space-based experiments: Discoverer \citep{1964JGR....69.3939S}, OGO1 \citep{1969ApJ...155..609C}, Pioneer8 \citep{1971JGR....76.1605L}, IMP 3, 4, 5, 6, 7 and 8 \citep{1965JGR....70.3515F,1966PhRvL..16..813F,1966JGR....71.1771B,1970ApJ...159...61H,1971ApJ...166..221H,1975ApJ...202..265G,1976ApJ...206..616M,1985ApJ...294..455B}, Voyager1-HET \citepads{1983ApJ...275..391W}, ISEE3-MEH \citepads{1986ApJ...303..816K}, CRRES \citepads{2000SoPh..195..175C}, EPHIN \citep{2016SoPh..291..965K}, AMS-01 \citepads{2000PhLB..490...27A,2000PhLB..494..193A}, PAMELA \citepads{2011Sci...332...69A,2013ApJ...765...91A,2016ApJ...818...68A}, and AMS-02 \citepads{2015PhRvL.114q1103A,2015PhRvL.115u1101A};

   \item Balloon-borne experiments: LEAP \citepads{1991ApJ...378..763S}, MASS 89 and 91\footnote{For MASS91, the low energy points (below the rigidity cutoff) are excluded from the fit.} \citepads{1991ApJ...380..230W,1999PhRvD..60e2002B}, SMILI-I and II \citepads{1993ApJ...413..268B,1995ICRC....2..630W}, IMAX92 \citepads{2000ApJ...533..281M}, CAPRICE 94 and 98 \citepads{1999ApJ...518..457B,2003APh....19..583B}, BESS93 to BESS-PolarII \citepads{2001AdSpR..26.1831S,2002ApJ...564..244W,2007APh....28..154S,2016ApJ...822...65A};

   \item Unnamed balloon-borne flights from the 50's through the 80's: \citetads{1956PhRv..104.1723M,1957PhRv..107.1386M,1959PhRv..116..462M,1957PMag....2..157F,1958Natur.181.1319F,1959PhRv..115..194M,1960JGR....65..767M,1961PhRv..121.1206A,1963PhRv..129.2275M,1964JGR....69.3097W,1964JGR....69.3293F,1964PhRv..133..818F,1964PhRvL..13..106O,1968JGR....73.4231O,1965JGR....70.2111F,1965JGR....70.5753F,1968JGR....73.4261F,1966JGR....71.1771B,1966P&SS...14..503C,1967PhRv..163.1327B,1967JGR....72.2765D,1967NCimA..47..189F,1967P&SS...15..715H,1971JGR....76.7445R,1972PhRvL..28..985R,1973ApJ...180..987S,1973ICRC....2..732G,1974JGR....79.4127R,1978ApJ...221.1110L,1979ICRC....1..330B,1983ApJ...275..391W,1987ICRC....1..325W}. 
\end{itemize}

\bibliography{nm_phi}

\end{document}